\documentclass[11pt]{article}
\usepackage{times}
\usepackage{sidecap}
\usepackage{geometry}
\usepackage{floatrow}
\usepackage{float}
\usepackage{hyperref}
\usepackage{graphics,graphicx}
\usepackage{subcaption}
\usepackage{epsf,wrapfig}

\usepackage{natbib,bm}

\usepackage{paralist}
\let\olditem\item
\renewenvironment{thebibliography}[1]{%
  \section*{\refname}
  \let\par\relax
  \renewcommand{\item}[1][]{\olditem[\textbullet]}%
  \inparaenum}{\endinparaenum}

\newcounter{aff}

\DeclareGraphicsExtensions{.jpg,.pdf,.png,.eps}

\usepackage{caption}
\usepackage{xcolor}

\usepackage{tcolorbox}

\bibpunct{(}{)}{;}{a}{}{,}

\usepackage{soul}

\definecolor{darkcyan}{rgb}{0.0, 0.0, 0.7}

\definecolor{boxcolor}{RGB}{193, 229, 255}
\definecolor{figboxcolor}{RGB}{193, 229, 255}
\definecolor{boxcolorgray}{RGB}{227, 227, 227}
\definecolor{darkcyan}{rgb}{0.0, 0.0, 0.7}

\sethlcolor{boxcolorgray}

\def\gtrsim{\lower 2pt \hbox{$\, \buildrel {\scriptstyle >}\over
{\scriptstyle \sim}\,$}}
\def\lesssim{\lower 2pt \hbox{$\, \buildrel {\scriptstyle <}\over
{\scriptstyle \sim}\,$}}

\def\approxlt{\lower.2em\hbox{$\buildrel < \over \sim$}}
\def\approxgt{\lower.2em\hbox{$\buildrel > \over \sim$}}

\def\be{\begin{equation}}
\def\ee{\end{equation}}
\def\bea{\begin{eqnarray*}}
\def\eea{\end{eqnarray*}}

\def\jref@jnl#1{{\rm#1}}
\def\aj{\jref@jnl{AJ}}			
\def\araa{\jref@jnl{ARA\&A}}		
\def\apj{\jref@jnl{ApJ}}			
\def\apjl{\jref@jnl{ApJ}}		
\def\apjs{\jref@jnl{ApJS}}		
\def\ao{\jref@jnl{Appl.~Opt.}}		
\def\apss{\jref@jnl{Ap\&SS}}		
\def\aap{\jref@jnl{A\&A}}		
\def\aapr{\jref@jnl{A\&A~Rev.}}		
\def\aaps{\jref@jnl{A\&AS}}		
\def\azh{\jref@jnl{AZh}}			
\def\baas{\jref@jnl{BAAS}}		
\def\jrasc{\jref@jnl{JRASC}}		
\def\memras{\jref@jnl{MmRAS}}		
\def\mnras{\jref@jnl{MNRAS}}		
\def\pra{\jref@jnl{Phys.~Rev.~A}}	
\def\prb{\jref@jnl{Phys.~Rev.~B}}	
\def\prc{\jref@jnl{Phys.~Rev.~C}}	
\def\prd{\jref@jnl{Phys.~Rev.~D}}	
\def\pre{\jref@jnl{Phys.~Rev.~E}}	
\def\prl{\jref@jnl{Phys.~Rev.~Lett.}}	
\def\pasp{\jref@jnl{PASP}}		
\def\pasj{\jref@jnl{PASJ}}		
\def\qjras{\jref@jnl{QJRAS}}		
\def\skytel{\jref@jnl{S\&T}}		
\def\solphys{\jref@jnl{Sol.~Phys.}}	
\def\sovast{\jref@jnl{Soviet~Ast.}}	
\def\ssr{\jref@jnl{Space~Sci.~Rev.}}	
\def\zap{\jref@jnl{ZAp}}			
\def\nat{\jref@jnl{Nature}}		
\def\iaucirc{\jref@jnl{IAU~Circ.}}
\def\aplett{\jref@jnl{Astrophys.~Lett.}}
\def\apspr{\jref@jnl{Astrophys.~Space~Phys.~Res.}}
\def\bain{\jref@jnl{Bull.~Astron.~Inst.~Netherlands}}
\def\fcp{\jref@jnl{Fund.~Cosmic~Phys.}}
\def\gca{\jref@jnl{Geochim.~Cosmochim.~Acta}}
\def\grl{\jref@jnl{Geophys.~Res.~Lett.}}
\def\jcp{\jref@jnl{J.~Chem.~Phys.}}	
\def\jgr{\jref@jnl{J.~Geophys.~Res.}}	
\def\jqsrt{\jref@jnl{J.~Quant.~Spec.~Radiat.~Transf.}}
\def\memsai{\jref@jnl{Mem.~Soc.~Astron.~Italiana}}
\def\nphysa{\jref@jnl{Nucl.~Phys.~A}}
\def\physrep{\jref@jnl{Phys.~Rep.}}
\def\physscr{\jref@jnl{Phys.~Scr}}
\def\planss{\jref@jnl{Planet.~Space~Sci.}}
\def\procspie{\jref@jnl{Proc.~SPIE}}

\def\arcdeg{\hbox{$^\circ$}}

\def\ion#1#2{#1$\;${\small\rm\@Roman{#2}}\relax}

\newcount\lecurrentfam
\def\LaTeX{\lecurrentfam=\the\fam \leavevmode L\raise.42ex
\hbox{$\fam\lecurrentfam\scriptstyle\kern-.3em A$}\kern-.15em\TeX}
\def\sizrpt{
(\fontname\the\font): em=\the\fontdimen6\font, ex=\the\fontdimen5\font
\typeout{
(\fontname\the\font): em=\the\fontdimen6\font, ex=\the\fontdimen5\font
}}
\definecolor{dark}{rgb}{ 0.34, 0.09, 0.49}

\usepackage{fancyhdr,url}
\usepackage{transparent}
\usepackage{tablefootnote}
\usepackage{mathpazo}

\begin{document}

\noindent
\begin{raggedright}
{\Huge\bfseries Tracking the flow of interstellar gas across scales: from galaxies to star clusters\\[0.5em]
\Large Unravelling the physics that regulate the evolution of the interstellar medium, and ultimately drive the formation of stars.
}\end{raggedright}\\

\noindent
Ana Duarte-Cabral$^1$, Maxime Valeille-Manet$^1$, Mathilde Bouvier$^{2,3}$, Miguel Querejeta$^{4}$, Jan Forbrich$^{5}$, Daniela Calzetti$^{6}$, Cecilia Bacchini$^7$, Frédérique Motte$^8$, Maria J. Jimenez-Donaire$^{9,4}$, Bruce G. Elmegreen$^{10}$, Dieu D. Nguyen$^{11}$, Alessio Traficante$^{12}$, Alvaro Hacar$^{13}$, Nicolas Peretto$^{1}$, Francesco Salvestrini$^{14}$, Laura Colzi$^{15}$, Thomas G. Williams$^{16}$, Timothy A. Davis$^{1}$, Tony Wong$^{17}$, Helena Faustino Vieira$^{18}$, Eva Durán-Camacho$^{19,20}$, Victor M. Rivilla$^{15}$, Kate Pattle$^{21}$, Sean T. Linden$^{22,23}$, Rosita Paladino$^{24}$, Ilaria Ruffa$^{25}$, Shih-Ping Lai$^{26}$, Elias Brinks$^{5}$, Eva Schinnerer$^{27}$, Drew Lapeer$^{28}$, Theotokis Georgatos$^{1}$, Miguel Pereira-Santaella$^{29}$, Ivanna Langan$^{15}$

\vspace{1em}
{
\footnotesize{
\noindent $^1$ Cardiff Hub for Astrophysics Research and Technology (CHART), School of Physical, Chemical and Environmental Sciences, Cardiff University, UK 
$^2$ Transdisciplinary Research Area (TRA) ‘Matter’ - Hertz Chair for Astrochemistry,  Universit\"{a}t Bonn, Am Propsthof\,49, 53121 Bonn, Germany 
$^3$ Leiden Observatory, Leiden University, P.O. Box 9513, 2300 RA Leiden, The Netherlands, 
$^4$ Observatorio Astronómico Nacional (IGN), Spain 
$^5$ University of Hertfordshire, UK 
$^6$ Department of Astronomy, University of Massachusetts, Amherst, MA, USA 
$^7$ DARK, Niels Bohr Institute, University of Copenhagen, Jagtvej 155, 2200 Copenhagen, Denmark 
$^8$ Univ. Grenoble Alpes, CNRS, IPAG, 38000 Grenoble, France 
$^9$ ESA/AURA, Space Telescope Science Institute, Baltimore, USA 
$^{10}$ Katonah, NY 10536, USA 
$^{11}$Department of Astronomy, University of Michigan, 1085 South University Avenue, Ann Arbor, MI 48109, USA 
$^{12}$ INAF Rome, Italy 
$^{13}$ University of Vienna, Austria 
$^{14}$ INAF Osservatorio Astronomico di Trieste, Italy 
$^{15}$ Centro de Astrobiología (CAB), CSIC-INTA, Carretera de Ajalvir km 4, Torrejón de Ardoz, 28850 Madrid, Spain 
$^{16}$ UK ALMA Regional Centre Node, Jodrell Bank Centre for Astrophysics, Department of Physics and Astronomy, The University of Manchester, Oxford Road, Manchester M13 9PL, UK
$^{17}$ Department of Astronomy, University of Illinois, Urbana IL 61801, USA 
$^{18}$ Department of Astronomy, Oskar Klein Centre, Stockholm University, AlbaNova University Centre, Stockholm, Sweden 
$^{19}$ Instituto de Astrof\'{i}sica de Canarias, V\'{i}a L\'{a}ctea S/N, E-38205 La Laguna, Spain
$^{20}$ Departamento de Astrof\'{i}sica, Universidad de La Laguna, E-38206 La Laguna, Spain
$^{21}$ Department of Physics and Astronomy, University College London, Gower Street, London WC1E 6BT, UK 
$^{22}$ School of Physics and Astronomy, Rochester Institute of Technology, 84 Lomb Memorial Drive, Rochester, NY 14623, USA
$^{23}$ Steward Observatory, University of Arizona, 933 N Cherry Avenue, Tucson, AZ 85721, USA 
$^{24}$ ESO Garching, Germany / INAF - IRA, Italy  
$^{25}$ INAF - Osservatorio Astrofisico di Arcetri / INAF - IRA, Italy 
$^{26}$ National Tsing Hua University, Taiwan 
$^{27}$ MPIA, Germany 
$^{28}$ University of Massachusetts Amherst, USA 
$^{29}$ Instituto de F\'isica Fundamental, CSIC, Calle Serrano 123, E-28006 Madrid, Spain
}
}
\begin{tcolorbox}[boxsep=3pt,left=0pt,right=0pt,top=1pt,bottom=1pt,colframe=boxcolor,colback=boxcolor]
\justify
\textbf{Abstract:} 
While the baryon cycle of the interstellar medium -- and its ability to form stars -- plays a key role in galaxy evolution, unravelling the physical processes driving its evolution remains a critical challenge, given that present-day observing facilities are unable to map entire galactic ecosystems down to the scales of individual star-forming clumps. Here, we outline the technical requirements that would enable, for the first time, to connect the small (pc) scale physics of star formation, to the large (tens-of-kpc) scales of galaxies, enabling an effective tracking of the flows of matter and energy across a vast range of environments, and answering the long-standing question of what regulates star formation across cosmic time.
\end{tcolorbox}

\newpage{}

\vspace{-0.7cm}
\section{Scientific context and open questions}
\label{sec:intro}

Much of the evolution of galaxies across cosmic time is driven by the formation and evolution of stars, thus tightly coupled to the physics of the interstellar medium (ISM). Made up of gas and dust, the ISM is a complex dynamic system in which the gas continuously transitions from a diffuse ionized medium, a warm atomic phase, and a much colder and denser molecular phase, a portion of which will ultimately fragment into star-forming cores. This cascade of energies and gas flows relies on the interplay of turbulence, gravity, feedback, and magnetic fields, all concurring within complex systems -- their host galaxies. 

Empirical relations often used in galaxy evolution models, such as the Schmidt-Kennicutt (SK) relation \citep[][]{Schmidt1959,Kennicutt1998} linking surface densities of gas and star formation rates (SFRs), or the existence of a universal stellar initial mass function \citep[e.g.][]{Kroupa2001,Chabrier2003}, intrinsically assume the ISM cycle has universal properties which lead to universal star formation (SF) relations.
However, the large-scale dynamics of galaxies (e.g. driven by differential rotation, bars, spiral arms, and interactions with external systems) create very different initial conditions from which molecular clouds form. In addition, the immense diversity of physical conditions found in those galaxies (in terms metallicities, radiation field, ISM pressure/density, turbulence, magnetic field strength, stellar feedback and active galactic nuclei (AGN) activity) directly impacts the way clouds evolve and fragment. It is thus perhaps not too surprising that we sometimes observe large variations of the shape of the stellar initial mass function \citep[e.g.][]{Schneider2018,Hosek2019} and of the gas-to-star conversion rates and efficiencies \citep[e.g.][]{Saintonge2011}. {\it This challenges the assumption of a Universal ISM-SF cycle}. However, a clear understanding of the physics driving any variations is still elusive. One example is the fact that we struggle to grasp what drives the formation of massive stars (and clusters), which then go on to play the most critical role in the baryon cycle of galaxies, through their driving of turbulence (with the injection of energy and momentum), and enrichment of the ISM with metals. 
Our inability to understand the conditions under which specific types of stars are able to form is mostly due to the current limitations when attempting to connect the vast range of scales involved: we can either study individual star-forming regions in detail, or study large galactic systems as a whole. However, to effectively measure cause and effect we need to probe the small-scale physics of the SF process \textit{and} capture the large-scale environment \textit{simultaneously}. This is incredibly challenging to do. 

Milky Way (MW) studies offer unparalleled physical resolution of the ISM and SF\footnote{See also the ALMA2040 white papers (WPs) by Sanchez-Monge et al., Kuffmeier et al.}, from the large statistics of surveys at pc-scales \citep[e.g.][]{Rigby2019,Duarte-Cabral2021}, to selected studies of individual clouds down to thousands of au scales \citep[e.g.][]{Hacar2018,Motte2022,Coletta2025}. Still, such studies sample only the limited range of environments available within our Galaxy, which do not reflect the diversity of environments in other galaxies. Additionally, placing individual clouds within their Galactic context remains challenging: uncertainties in distances and in the structure of the MW often make it difficult to establish their precise location relative to spiral arms, the bar, and the Galactic centre, hampering our ability to connect SF to the large-scale galactic environment. Furthermore, while there is a substantial population of clouds in the Galaxy forming high-mass stars \citep[e.g.][]{Molinari2025}, there are no candidate regions that could be progenitors of super star clusters, SSCs\footnote{Defined as compact (R $\sim 1 - 5$\,pc), young (1 -- 100 Myr) and extremely massive (M$_*> 10^4$M$_\odot$) stellar clusters.} identified anywhere outside the Galactic centre \citep[e.g.][]{Longmore2014}. This limits our ability to use Galactic studies alone to fully understand and extrapolate the physical conditions that enable the formation of the most massive clusters observed in the Universe. The ability to do so thus rests on our nearby Universe.

ALMA and JWST have started to lead the way into more in-depth studies of the ISM and SF in nearby galaxies, allowing systematic characterization of global cloud properties, scaling relations, and emerging stellar clusters \citep[e.g.][]{Davis2022,Schinnerer2024,Leroy2025,Knutas2025}. Even so, most studies focus on the bright, metal-rich star-forming nearby galaxies, and the ISM cycle in faint dwarf/metal-poor systems remains widely unexplored due to the challenging nature of finding suitable and bright molecular tracers, beyond the most nearby systems \citep[e.g.][]{Hunter2024}. In addition, even the most comprehensive high-resolution studies of nearby star-forming galaxies only offer resolutions of $\sim$10-100\,pc, and the internal physics of the cold molecular gas remains largely inferred rather than directly measured. In other words, in external galaxies, {\it we may know where the molecular gas is, but we do not know how it might be forming stars}, as we do not reach the critical fragmentation scales of clumps -- the self-gravity-dominated pc-sized reservoirs of dense gas, which are the direct progenitors of individual stellar clusters.

Numerical models have since long helped in our understanding of how the gas and SF respond to specific changes in the environmental conditions \citep[e.g.][]{Klessen2005,Wurster2018,Seifried2020}. But just as for observations, they typically focus on specific scales: either the modelling of individual clouds (with loss on the larger scale context), or on the modelling of kpc-sized regions/entire galaxies (with loss on the smaller scale physics, requiring the use of sub-grid models for the unresolved SF process). Connecting the many orders of magnitude range in scales and densities involved in the ISM-SF cycle is prohibitively computationally expensive, with most attempts having to resort to ''zoom-in'' techniques -- where only a sub-set of a larger-scale simulation is re-simulated at higher resolution \citep[e.g.][]{Grand2017,Smith2020,Cusack2026}, and whose results are then inevitably limited on number statistics.

The challenging nature of simultaneously probing deep inside clouds and retaining the large-scale context, prevents us from disentangling the response of the ISM to environmental changes. We thus remain with the question of {\it What physical processes regulate the evolution of the ISM, and ultimately drive the formation of stars?} This question is central to grasp galaxy evolution, but answering it requires the ability to {\bf systematically connect the gas cycle across the vast scales of a galactic ecosystem (from tens-of-kpc down to pc-scales)}, that no current observatory is capable of.\\

Technically, the maximum baselines of ALMA -- the largest current (sub)mm interferometric facility in the world -- already allow reaching physical resolutions of $\sim$1-5\,pc in most bands, for galaxies at distances up to 10\,Mpc. But in practice, the lack of sensitivity and image fidelity make the pc-scale resolution mapping of entire galaxy systems, where faint (and often extended) emission dominates, unfeasible\footnote{As a reference, for a 3$\sigma$ continuum detection at 230\,GHz of a typical high-mass star-forming clump (with $\sim$2000\,M$_{\odot}$ within $R\sim$2.5\,pc) at the distance of NGC628 (9.8\,Mpc), ALMA would require $\sim$88\,days ($\sim$22\,days for ALMA-WSU) per pointing (with 4.7\,pc resolution, and a field of view, FoV, of $\sim$1.5\,kpc$^{2}$). For spectral lines, reaching the required sensitivity would take $\sim$50\,days with ALMA -- see Sect.\,\ref{sec:simulations} for more details.} for galaxies beyond the Local Group ($d>$1\,Mpc). Within the local 1\,Mpc, ALMA can reach the required sensitivity at a few pc resolution, but in practice such studies would suffer from poor image quality, cover only relatively small areas\footnote{For example, if we take M33 (at 0.9\,Mpc), we could achieve the same sensitivities as for the NGC628 example above, with $\sim$8\,min/pointing. However, each pointing would have a FoV $\sim$0.01\,kpc$^{2}$, thus to map the entire galaxy would take $>$100 days of observations. Covering just an area of 1.5\,kpc$^{2}$ (similar to that covered on a single pointing of NGC628) would require 13h of observations, which is technically within reach, but such short integration times per pointing would result in very poor uv-coverage, and thus unreliable line emission maps. If pushing to higher resolutions (e.g. 2\,pc), ALMA would require 3\,h/pointing to reach similar line sensitivities, thus to simply cover an area of 1.5\,kpc$^{2}$ (in just one local galaxy) would take 12 days - which is at the very limit of what ALMA would be able to observe.}, and be inevitably restricted to the range of physical conditions available within the few galaxies in the Local Group (i.e. 3 main spiral galaxies, and a few tens of dwarfs and ultra-faint dwarfs).

\begin{figure}
\centering
\includegraphics[width=\linewidth]{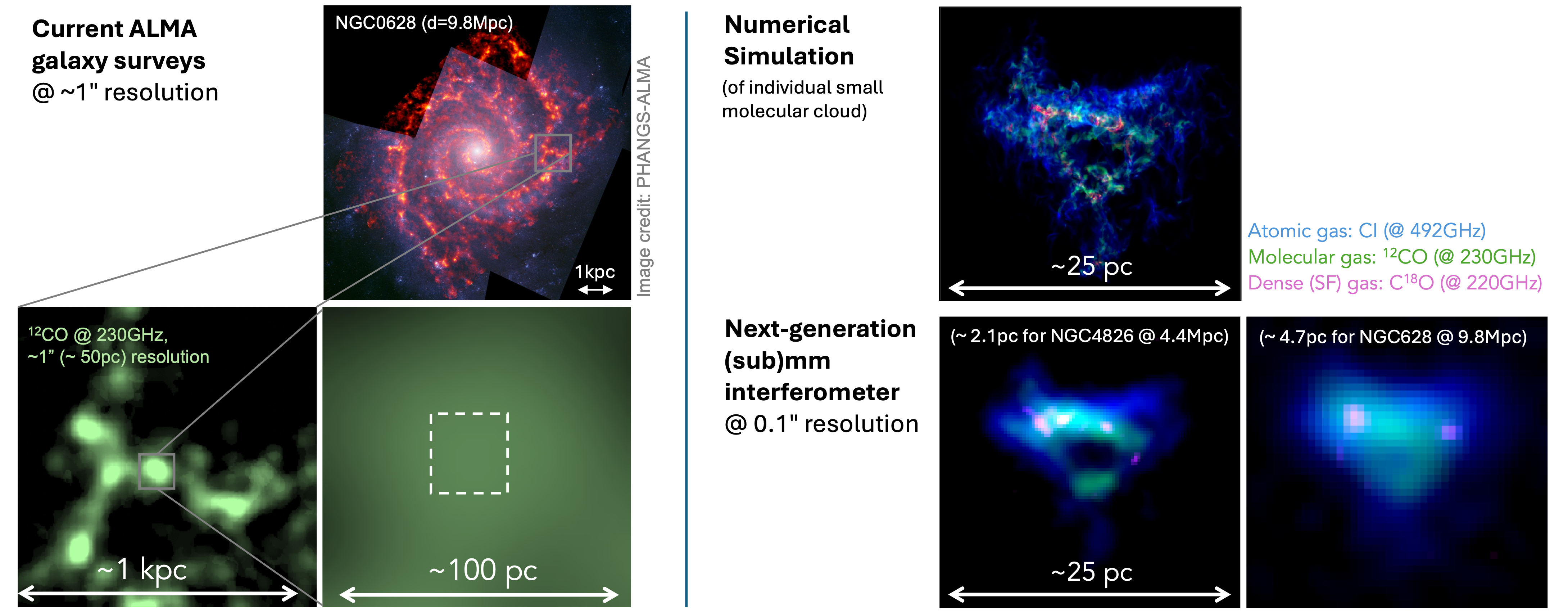}
\caption{\small {\textit{Left:} PHANGS HST/CO emission of NGC628 (top) and a set of zoom-ins (bottom) on a $\sim$1\,kpc and $\sim$100\,pc patch of the CO emission, illustrating the level of detail of current molecular gas observations on tens-of-pc scales (i.e. 1''). Dashed box on bottom-right panel shows a 25\,pc region equivalent to the size of the right panels. \textit{Right:} Prospect for ALMA2040's ability to detect and resolve the emission at pc scales using multiple gas tracers. Colours show the integrated intensity of C{\sc i} (blue), $^{12}$CO (green) and C$^{18}$O (magenta), from synthetic observations of a molecular cloud simulation \citep[][]{Priestley2023}, at its native resolution and with no noise (top), as well as placed at the distance of two example galaxies (NGC4826 at 4.4\,Mpc and NGC628 at 9.8\,Mpc, bottom left and right resp.), with a resolution of 0.1'' (resolving 2.1\, and 4.7\,pc, resp.) and sensitivity achievable with ALMA2040 -- see Sect.\,\ref{sec:simulations} for more details.
\vspace{-0.3cm}}}
\label{fig_zoomNGC628}
\vspace{-0.2cm}

\end{figure}

A true step-change to grasping the ISM-SF cycle requires a next-generation (sub)mm facility capable of mapping emission in multiple wavelength bands, across large areas of galaxies and down to the pc-scales of star-forming clumps (see Sect.\,\ref{sec:tech_requirements}, and Fig.\,\ref{fig_zoomNGC628}). The prospect of a facility with 10x higher sensitivity, and unprecedented image-fidelity recovering all angular scales, would finally allow connecting the kpc-to-pc-scale physics of the SF process for targets up to 10\,Mpc away. This would give us access to hundreds of galaxies spanning the full range of masses, morphologies, metallicities and nuclear activities, turning the connection between pc-scale physics and galactic environment into a statistical measurement rather than a collection of case studies from the Local Group. Here we highlight some of the key science that would be enabled by such a facility.

\subsection{Tracking the assembly of molecular gas}
\label{sec:assembly}

The assembly of gas into cold and dense molecular clouds is the first step of the SF process, as it defines the mass build up -- the reservoir and replenishing rates -- of star-forming gas. Molecular gas is formed from the convergence of the diffuse ISM, through compressive shocks: e.g. from changes in the galactic potential, shear, cloud collisions, feedback, or smaller-scale shocks from supersonic turbulence. Linking the SF-potential of clouds to their assembly history requires simultaneously probing the kpc-scale environmental properties within the host galaxy, the gas phase transition between the atomic and molecular gas at $\sim$10-100\,pc scales, and the gas flows into clumps at pc-scales. To achieve this, a next generation observatory must offer efficient high-sensitivity and high-resolution mapping of species across multiple bands, tracing the different phases of the ISM (Fig.\,\ref{fig_zoomNGC628}), such as the atomic/warm molecular gas (with C{\sc{i}}); the cold molecular gas (e.g. with CO isotopologues, HCN, HCO$^{+}$, N$_{2}$H$^{+}$); and shock tracers related to the cloud assembly process \citep[with HNCO, CH$_{3}$OH and SiO, e.g.][]{Jimenez-Serra2010,Bogner2025}. 

Accretion rates across these phase transitions can be inferred from analysis of the gas kinematics. Full-galaxy spectral-line mapping yield velocity fields with circular and non-circular components, which can be separated by harmonic decomposition or forward kinematic modelling of the data cubes \citep[e.g.][]{Wong2004,Oh2019,DiTeodoro2021}\textcolor{blue}. The residual motions, combined with the gas surface density, give mass-flow rates as a function of scale: from bar- and spiral-driven radial flows at kpc scales, through convergence of flows at the cloud--clump interface ($\dot{M} \sim \Sigma_{\rm gas}\, v_{\rm in}\, \ell$\footnote{Measurable with some assumptions on the projected geometry of the system.}), down to the pc-scale infall signatures of individual clumps \citep[e.g.][]{Traficante2017,Morii2026}. Applied uniformly across a galaxy sample, this could provide the assembly-rate function of star-forming structures in different environments. 
In addition, in nuclear regions, this could also allow measuring the gas flows potentially fuelling the central black hole, as well as determining how the cold gas properties, dense-gas fractions and SF efficiencies respond as a function of distance from an active nucleus, thus linking the SF cycle to AGN activity and black hole growth within the same dataset.
Moreover, this wealth of spectral line data would provide key insights into how the chemical tracers typically used for the characterisation of the local physical conditions, change with different environments (e.g. with varying metallicity, ionising radiation level, cosmic ray flux, shear rates, thermal pressure). 

ALMA2040's unique prospect to detect and resolve the cold component of the ISM gas flows (from kpc down to pc-scales, on a statistically significant sample of galaxies and environments), would have strong synergies with SKAO, JWST, Roman (and potentially future HWO), namely with those observatories' ability to capture the distribution of the more diffuse material. This includes the HI emission from the neutral atomic medium, the optical extinction probing the distribution of dust \citep[e.g.][]{FaustinoVieira2023}, the mid-IR observations of warm (shock-heated) H$_{2}$ \citep[e.g.][]{Lopez2025}, and the emission from polycyclic aromatic hydrocarbons (PAH). In fact, PAH emission observed in nearby galaxies with the JWST is already showcasing exquisitely intricate structures of the ISM at a few tens-of-pc scales \citep[e.g][]{Sandstrom2023, Dale2025, Rodriguez2025}. However, PAH emission is notoriously hard to interpret and the physical properties of the material it traces remain poorly constrained \citep[e.g.][]{Chastenet2023, Pedrini2024}. Our full understanding of the ISM thus relies on the combined information we can get from multiple gas tracers and thermal continuum emission.

\subsection{Fragmentation and kinematics of clouds}
\label{sec:fragmentation}

To infer how SF is affected by a cloud's assembly history and environmental conditions (Sect.\,\ref{sec:assembly}), it is imperative to track the fragmentation of clouds down to pc-scales (the critical scales providing the mass reservoir of individual stellar clusters). This can be achieved by mapping the cold dust emission (in the 1-3\,mm wavelength range) tracing the population of star-forming dense clumps, as well as the slightly hotter dust (in higher frequency bands) crucial for lower metallicity environments and photon-dominated regions. These capabilities are currently out of reach: the continuum sensitivity of current and planned facilities severely limits our ability to detect dust emission at resolved cloud scales even in the most nearby galaxies beyond the Magellanic Clouds. With dust being less dependent on the chemical evolution of the ISM than molecular gas\footnote{Molecular gas emission depends not only on the local conditions such as density and temperature (similarly to the dust emission), but is also subject to changes in the relative abundance of specific species.}, the ability to detect it would provide a reliable tracer of the ISM mass distribution\footnote{The reference high-mass star-forming clump given in Sect.\,\ref{sec:intro} that would take $\sim88$days of observations with ALMA, would become feasible in just $\sim1.3$\,h with ALMA2040 (assuming 40$\times$ better continuum sensitivity).}. Comparisons of spectral-line and dust measurements can further provide accurate molecular abundances, improving our understanding of ISM chemistry across different environments \citep[e.g.][see also the ALMA2040 WPs by den Brok et al. and Bouvier et al.]{Elmegreen2013}. This is particularly critical in more unusual environments, including galaxies with lower metallicities where there are lower dust abundances, and where the chemical pathways to form the typical gas tracers we use in the MW are not necessarily the same \citep[e.g.][]{Rubio2015,Archer2025}.

\begin{figure*}
\centering
\includegraphics[width=0.9\linewidth]{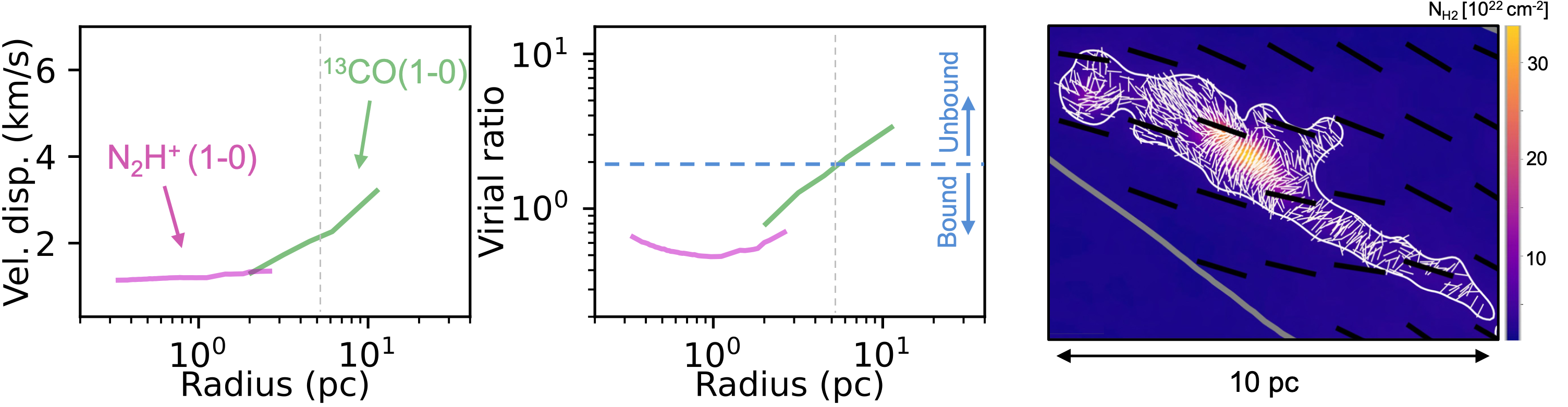}
\caption{\small{Left/centre: Profile of a high-mass star-forming cloud (SDC35.429+0.138) in the MW \citep[adapted from][who studied a sample of 27 IRDCs, all showing similar behaviours]{Peretto2023}, showing the velocity dispersion (left) and virial ratio (centre), as a function of radius from the centre of the clump, using N$_2$H$^+$ (purple) as a dense-gas tracer, and $^{13}$CO (green) as an intermediate-density tracer. This demonstrates the dynamical decoupling of the clump at a few pc scales (vertical dashed lines), with a flattening of the velocity dispersion at small radii, and the higher stability of the cloud at larger scales. Right: Example of the change in the magnetic field morphology from cloud to clump scales from another MW high-mass star-forming cloud \citep[SDC34.370, adapted from][]{Ramkumar2026}. Lines show the magnetic field direction from Planck (at $\sim$5\,pc resolution, black), and JCMT-POL2 (at sub-pc resolution, white).
\vspace{-0.3cm}}}
\label{fig_decoupling}
\vspace{-0.3cm}
\end{figure*}

Such dust fragmentation properties can be connected to the kinematics of the hierarchical gas flows derived from molecular lines tracing the intermediate densities of molecular clouds (e.g. $^{12}$CO, $^{13}$CO), and the cold dense gas (e.g. C$^{18}$O, HCN, HNC, N$_2$H$^+$, HCO$^+$, DCN). Typical studies of such flows are confined to selected samples of molecular clouds in the MW, especially for dense gas tracers where their fainter emission make them impossible to map at pc-scales in nearby galaxies, with enough spectral resolution to resolve the velocity structure \citep[observations from e.g.][have mapped the integrated intensity of dense gas tracers in nearby galaxies, but at hundreds-of-pc to kpc scales and with velocity resolutions of 4-10\,km\,s$^{-1}$, thus unable to resolve the pc-scale internal motions of the gas of the order of $\sim$1-2\,km\,s$^{-1}$]{Jimenez-Donaire2019,Stuber2025}. Recent MW studies of the hierarchy of the kinematics within molecular clouds have suggested that -- at least in high-mass star-forming regions -- the pc-scale clumps have more kinetic energy than expected from a turbulent cascade \citep[e.g.][]{Traficante2020}, and seem dynamically decoupled from their parent cloud  \citep[e.g.][Fig.\,\ref{fig_decoupling}]{Peretto2023}. These motions have been interpreted as driven by the gravitational collapse of the clump structures within more stable clouds, but these results arise from small samples of high-mass star-forming regions in typical MW-disc conditions. 
A true understanding of the physics driving the flows of gas and cloud fragmentation requires studying this process elsewhere in the Universe, by probing the critical fragmentation and kinematics of the cold gas that is actively forming stars in a wide sample of galaxies in the local Universe -- a task that only a next-generation (sub)mm facility can hope to achieve.

\subsection{Tracing the deeply embedded/early stages of star formation}
\label{sec:SFtracers}

A long-standing unknown of the SF process is the physical driver of the low SFE of clouds, and whether it changes with environment. Indeed it is incredibly hard to discern whether an observed change in SFR is fuel- or efficiency- driven \citep[e.g.][]{Ellison2021}, but even more so, to link those changes to the environmental conditions of the clouds actively forming stars. To add to the confusion, the various ways in which we measure SFR and SFE in the MW and nearby galaxies are not always easily comparable \citep[e.g.][]{Leroy2008,Calzetti2013,Barnes2017}, and projection effects can also affect these measurements \citep[e.g.][]{Bacchini2019}. 
Additionally, extragalactic studies are often unable to probe the deeply embedded population of forming stars, which are IR-dark. This phase is only seen with far-IR emission (for which there is no current/planned facility able to resolve clump-scales anywhere outside our Galaxy) or through tracers of the early feedback from the forming stars -- before clusters are able to ionise and clear their surrounding gas. This makes it difficult to directly correlate the existence of dense gas, to its on-going SF activity. 

A future observatory capable of resolved multi-line and continuum observations of the ISM across galaxies down to star-forming clump scales at exquisitely high sensitivity (Sect.\,\ref{sec:fragmentation}), would be uniquely able to provide reliable measures of the on-going SF, still embedded in clouds, by tracing their early protostellar feedback signatures. This could be achieved for targetted observations, especially for the most massive protoclusters, through chemical probes of the on-going SF in the (sub)mm regime (see also the ALMA2040 WP by Sun et al.). These include tracers of hot cores \citep[such as methanol, or vibrationally excited HC$_{3}$N, e.g. ][]{Rico-Villas2021, Harada2024}; protostellar outflow shock tracers \citep[such as high-velocity SiO or CO, which can provide indirect measures of the accretion rate and evolutionary stage of the forming clusters, e.g.][]{Duarte-Cabral2013, Maud2015}; and H{\sc{ii}} regions tracing the ionised gas around young massive stars \citep[with Hydrogen recombination lines such as H41$\alpha$, H30$\alpha$ or H26$\alpha$, e.g.][]{Baez-Rubio2018}. Quantifying the level of feedback from young stellar clusters (and how it changes with cluster mass, and time) is also critical to understand the wider role played by stellar feedback in dispersing molecular clouds, a crucial step for the UV radiation from the massive stars to escape into the ISM, and with potential consequences for regulating the SFE and SFR of galaxies.

Combining the feedback signatures from the deeply embedded population of clusters (still IR-dark) with observations in the optical/near-IR (tracing the emerged cluster population), would provide a more complete picture of the SF history (and more accurate measures of the SFR and SFE) of individual clouds. This could also potentially allow the calibration of the current methods to infer those for more distant unresolved galaxies. This characterisation via spectral lines has strong synergies with current JWST/HST observations \citep[of young and partially embedded stellar clusters, e.g.][]{Sun2024}, but also with the upcoming ELT (tracing the near-IR emission from embedded clusters at mas resolutions) and ngVLA observations (tracing ionised gas and H{\sc ii} regions), as well as with any future far-IR facility capable of resolving IR emission from the forming protostars. Linking SF variations to the environment will finally reveal the physical drivers of empirical relations such as the SK relation (and thus the origin of observed SF rates and efficiencies).

\subsection{The role of magnetic fields in the ISM-SF cycle}
\label{sec:Bfields}

Magnetic fields are believed to play a key role in channelling gas from the diffuse medium into dense star-forming cores, potentially regulating the gravitational collapse of clouds \citep[e.g.][]{Koch2012}. Yet, observational studies in nearby galaxies suggest that large-scale magnetic fields are largely dragged by the large-scale motions of the galaxy \citep[e.g.][]{Borlaff2021} -- that they ``go with the flow'', rather than dictate it.  Moreover, both supernova and nuclear feedback have been shown to reorganise magnetic fields on galactic scales \citep[e.g.][]{Pattle2021,Lopez-Rodriguez2025}. It therefore remains uncertain under which physical conditions and on which spatial scales magnetic fields significantly influence the structure and evolution of the ISM, provide appreciable support against gravitational contraction, or affect the fragmentation of clouds into gravitationally bound (and potentially dynamically decoupled) collapsing clumps.

Recent results towards nearby MW star-forming regions suggest that there might be a change in magnetic field morphology from cloud to clump-scales, which would imply a change in the relative importance of the magnetic field support against gravity, with magnetic fields becoming sub-dominant on clump-scales \citep[e.g.][and Fig.\,\ref{fig_decoupling}]{Ramkumar2026}. Results from \citet{Lee2025} further suggest that the fragmentation properties of clumps into cores depends on the properties of the magnetic fields on clump scales, thus anchoring the importance of capturing this critical regime. However, magnetic field observations remain technically challenging due to the low percentages of polarised emission. Furthermore, magnetic field measurements are drawn either from plane of the sky dust polarisation measurements and then rely on a number of techniques to infer the magnetic field strengths \citep[e.g.][]{Davis1951,Chandrasekhar1953,Koch2012}; or from measurements of the Zeeman splitting \citep[which are very hard to obtain, e.g.][]{Crutcher2010}. The velocity gradient technique allows to measure magnetic fields without the use of polarised emission, but it relies on the assumption that the kinematics of the gas is governed by a turbulent magnetised medium \citep[rather than gravity-dominated motions, e.g.][]{Gonzalez-Casanova2017}. Molecular line polarisation (e.g. of CO through the Goldreich-Kylafis effect) is another avenue for measuring magnetic fields, but it requires extremely high-sensitivity observations and its interpretation is non-trivial \citep[e.g.][]{Cortes2021}. This challenging nature of magnetic field observations means that systematic comparisons across spatial scales and physical environments remain limited (see also the Magnetic Universe WP by Girart et al.). 

The prospect of ALMA2040's high-sensitivity polarisation capabilities for both the continuum and lines would make it possible to map magnetic fields in nearby galaxies, taking advantage of the multiple techniques mentioned above as most suitable for different scales and density regimes. This would have synergies with SKAO which can provide complementary measurements of the magnetic field strength through Zeeman observations of H{\sc i} and OH \citep{Bourke2026}. This would enable unprecedented progress in tracking the evolution of magnetic fields across scales, test how they might vary with environment, and understand at what physical scale magnetic fields transition from being largely advected by galactic gas flows to dynamically influencing cloud collapse and fragmentation, ultimately affecting the end result of the SF process.

\section{ALMA2040: Technical Requirements}
\label{sec:tech_requirements}

This science requires a next-generation (sub)mm interferometer with technical requirements well beyond simple upgrades to the current facilities. For instance, the science presented in Sect.\,\ref{sec:assembly}, \ref{sec:fragmentation} and \ref{sec:SFtracers} requires a least a $10\times$ (ideally $20\times$) increase in instantaneous spectral line sensitivity compared to ALMA ($\sim$8-16$\times$ compared to anticipated ALMA-WSU); and a $40\times$ increase in continuum sensitivity compared to ALMA ($\sim$20$\times$ compared to anticipated ALMA-WSU). The science in Sec.\,\ref{sec:Bfields} also requires full polarisation capabilities with polarisation purity of  0.1\%. Besides these sensitivity requirements, all of the above science also require fast mapping speeds for large mosaics; access to large angular scales; and good image fidelity - as they all rely on the ability to connect the vast scales of the ISM, from kpc- down to pc-scales.

These advances are expected to be achievable with ALMA2040 through innovative developments in receiver technology such as focal plane arrays, simultaneous multi-band capabilities, and ultra-wide bandwidths, and substantially increasing the number of antennas (and thus collecting area), providing also excellent image fidelity across a range of angular scales (through exquisite uv-coverage, and potential combination with larger single dish telescope(s) of $\sim$20-30\,m for recovery of the total power and larger angular scales). Furthermore, an increase of the maximum baseline to $\sim$30\,km achievable within the Atacama astronomical park (2x larger than ALMA) would allow resolving 1\,pc at distances of $\sim$20\,Mpc. The critical pc-scales of clump decoupling (Fig.~\ref{fig_decoupling}) is therefore accessible not merely for the nearest systems but for several hundreds of galaxies, with a wide range of masses, metallicities, galaxy types, and nuclear activities. By enabling an effective tracking of the kinematics and chemical evolution of interstellar gas through the hierarchical structures of the ISM down to pc-scales across such a vast range of environments, this next-generation facility will finally allow us to answer the long-standing question of what regulates SF in galaxies.

\subsection{Simulations of expected sensitivity/feasibility}
\label{sec:simulations}

In order to simulate what ALMA2040 with the above technical capabilities would be able to achieve, we have used a numerical simulation of a relatively small molecular cloud formed through cloud collision of two spherical clouds of $10^{4}$\,M$_{\odot}$ within 19\,pc radius (each), colliding with a relative velocity of 14\,km\,s$^{-1}$ \citep[from][]{Priestley2023}. Here we place the observer such that the cloud is viewed with a 45$\arcdeg$ inclination with respect to the collision axis, and ran it through radiative transfer using RADMC-3D \citep[][]{Dullemond2012} to obtain the datacubes of synthetic observations for C{\sc i} (at 492\,GHz), as well as $^{12}$CO and C$^{18}$O emission (at 230\,GHz and 220\,GHz respectively). We mimicked the ALMA2040 observations, by placing the cloud at the distance of NGC4826 (i.e. 4.4\,Mpc) and NGC628 (i.e. 9.8\,Mpc), and assuming a representative resolution of 0.1'' (i.e. 2.1\,pc and 4.7\,pc respectively). We then computed the sensitivity achievable using the ALMA sensitivity calculator, assuming the declination of NGC628, and a 10x increase in line sensitivity compared to ALMA (i.e. a factor of 100 in observing time). The top-right panel of Fig.\,\ref{fig_zoomNGC628} shows the original integrated intensity from the synthetic observations, prior to beam convolution or noise injection. The two bottom-right panels show what ALMA2040 could observe for the two galaxies, with an rms noise level of 0.24/0.26/0.34\,K in 0.2\,km\,s$^{-1}$ channels for $^{12}$CO/C$^{18}$O/C{\sc i} respectively, which would be achievable in 12h of observations. As a reference, the current ALMA 12m array would not only require $\sim$50 days to achieve the same sensitivity at this resolution, but those observations would lack severely in image fidelity and ability to recover flux at all scales. With the planned ALMA WSU upgrade, even if line sensitivity might improve marginally, it will not reach the required levels to make these observations feasible, and it will provide no improvement on the instant uv-coverage, leaving the lack of image fidelity as a major drawback for these types of large-scale mapping observations. Furthermore, ALMA2040's ability to observe \textit{all bands} simultaneously would substantially optimise the observing strategies and time usage that would be needed for capturing the multiphase nature of the ISM which inherently requires multi-band observations. Note, that Fig.\,\ref{fig_zoomNGC628} only shows a small region of 25\,pc in size, as that is the size of the simulated molecular cloud, but the field of view of the ALMA2040 would be of the order of 25-30'' at $\sim$230\,GHz (for antennas of 12-15\,m diametre), thus capturing regions of $\sim1.5$\,kpc$^2$ per pointing for a galaxy like NGC628 (i.e. similar to the region illustrated in the bottom-left panel of Fig. \,\ref{fig_zoomNGC628}). 

It is worth noting that the simulation shown here is of a relatively low-mass star-forming region, and not representative of a typical high-mass star/cluster forming region (examples of such regions as observed in the Galaxy easily comprise $\sim10^{5}$\,M$_{\odot}$ within 30\,pc). These line emission brightnesses should thus be taken as a relatively conservative estimate for active star-forming disc galaxies like our own, with relatively high metallicities, especially for the dense-gas tracers, which this simulation has particularly little of. Nevertheless, one should also consider that high-mass star-forming regions in lower metallicity environments, with much lower abundances of these molecular species, might struggle to achieve these levels of surface brightness, thus possibility requiring more targetted observations and/or of more nearby systems. These could be potentially be selected from lower-resolution surveys of entire galaxies mapping higher density tracers at 50-100\,pc scales \citep[e.g.][]{Stuber2025}, that ALMA2040 would make possible for hundreds of galaxies (see the ALMA2040 WP by den Brok et al.), making this science truly within reach with a powerful next-generation (sub)mm facility such as that proposed with ALMA2040.

\section*{Acknowledgements}
This work was developed as part of the broader ALMA2040 community effort. We acknowledge the Lorentz Center and NOVA for supporting the community workshops and coordination activities that helped initiate and shape the ALMA2040 White Paper process. We also thank the reviewers, P. Hennebelle and E. Emsellem, for their useful feedback which helped the clarity of the WP. ADC acknowledges the support from a Royal Society University Research Fellowship (URF/R1/19160 and URF/R/241028). MVM and NP are supported by the Science and Technology Facility Council Small Award APP30146. MB acknowledges the support from the European Research Council Advanced Grant MOPPEX 833460. EDC acknowledges the support from the European Union (Widening Participation, ExGal-Twin, GA 101158446). SPL acknowledges support from the National Science and Technology Council of Taiwan (grants NSTC 115-2112-M-007-017-MY3 and NSTC 115-2918-I-007-013).\\

\bibliographystyle{apj_mod}
{\bibliography{references}}
\end{document}